\documentclass[twocolumn,pre,nofootinbib,floatfix,superscriptaddress,showkeys]{revtex4-1}

\usepackage{hyperref}
\usepackage{dcolumn}
\usepackage{graphicx}
\usepackage{rotating}
\usepackage{amsmath,amsfonts,amssymb}
\usepackage{multirow}

\begin{document}
\title{Gender, Productivity, and Prestige in Computer Science Faculty Hiring Networks}

\author{Samuel F. Way}
\email{samuel.way@colorado.edu}
\affiliation{Department of Computer Science, University of Colorado, Boulder CO, 80309 USA}

\author{Daniel B. Larremore}
\email{larremore@santafe.edu}
\affiliation{Santa Fe Institute, Santa Fe NM, 87501 USA}

\author{Aaron Clauset}
\email{aaron.clauset@colorado.edu}
\affiliation{Department of Computer Science, University of Colorado, Boulder CO, 80309 USA}
\affiliation{BioFrontiers Institute, University of Colorado, Boulder CO, 80303 USA}
\affiliation{Santa Fe Institute, Santa Fe NM, 87501 USA}

\begin{abstract}
Women are dramatically underrepresented in computer science at all levels in academia and account for just 15\% of tenure-track faculty. Understanding the causes of this gender imbalance would inform both policies intended to rectify it and employment decisions by departments and individuals. Progress in this direction, however, is complicated by the complexity and decentralized nature of faculty hiring and the non-independence of hires. Using comprehensive data on both hiring outcomes and scholarly productivity for 2659 tenure-track faculty across 205 Ph.D.-granting departments in North America, we investigate the multi-dimensional nature of gender inequality in computer science faculty hiring through a network model of the hiring process. Overall, we find that hiring outcomes are most directly affected by (i) the relative prestige between hiring and placing institutions and (ii) the scholarly productivity of the candidates. After including these, and other features, the addition of gender did not significantly reduce modeling error. However, gender differences do exist, e.g., in scholarly productivity, postdoctoral training rates, and in career movements up the rankings of universities, suggesting that the effects of gender are indirectly incorporated into hiring decisions through gender's covariates. Furthermore, we find evidence that more highly ranked departments recruit female faculty at higher than expected rates, which appears to inhibit similar efforts by lower ranked departments. These findings illustrate the subtle nature of gender inequality in faculty hiring networks and provide new insights to the underrepresentation of women in computer science.
\end{abstract}


\maketitle

\section{Introduction}
Women continue to be dramatically underrepresented in computer science, receiving only 18\% of bachelors' degrees and 20\% of doctorates in 2011,\footnote{\small{\url{http://nces.ed.gov/programs/digest/2013menu_tables.asp}}} and are estimated to hold fewer than 20\% of technical positions in the computing industry.\footnote{\small{\url{ http://cnet.co/1GZh268 }}} Women are especially underrepresented in the professoriate, making up only 15\% of tenured or tenure-track faculty in computer science departments~\cite{clauset2015systematic}. Understanding the causes of gender imbalance in faculty hiring would illuminate the underlying social processes that shape academic disciplines, and facilitate efforts both to support equal opportunities and to address the many non-meritocratic differences in male and female faculty experiences~\cite{ellemers:2014,katz:etal:2014,shen:2013}. These differences include disparities in tenure rates, competency evaluations, remuneration, allocation of research facilities, and grant competitions. Rectifying these differences and improving the gender balance in computer science would serve not only to advance social justice but would also promote the sort of diversity in skills and research approaches that has been found to improve group performance~\cite{page2008}, particularly in innovation-focused industries \cite{kets2015challenging}.

Much of the past research on gender imbalance among faculty has focused on the ``leaky pipeline,'' the name given to the observation that women leave science, technology, engineering and mathematics (STEM) fields at greater rates than men at every stage of an academic career, from grade school to full professor~\cite{hill2010so}. At the faculty hiring stage of the pipeline, several experimental studies have aimed to identify the causes of gender imbalance~\cite{moss2012science,ceci2014women,williams2015national}. However, these have yielded inconsistent, even contradictory findings, and little past work has focused specifically on computer science.

Essentially, faculty hiring is a community-based competitive process of subjective expert evaluations under conflicting and evolving preferences; that is to say, it's complicated. These features, along with the non-independent nature of hiring outcomes, make it difficult to reliably assess the presence and source of real biases. Here we investigate the role of gender in faculty hiring in computer science using a novel network model of the hiring process itself, across institutions and time. We then use this model to study the hiring histories of individual institutions and the experiences of individual faculty. We train this model using comprehensive data on the hiring outcomes, scholarly productivity, and gender of 2659 tenured or tenure-track faculty across all 205 computer science Ph.D.-granting departments in the United States and Canada~\cite{clauset2015systematic}.

Many studies have found evidence of gender bias in academia. 
For instance, male faculty in the life sciences tend to train fewer female graduate students and postdocs, relative to female representation in the pool of trainees~\cite{sheltzer2014elite}. This tendency is more pronounced at elite institutions, which tend to produce the majority of future faculty~\cite{clauset2015systematic}. Women often perceive greater barriers to becoming faculty than do men~\cite{van2004academic}, which may discourage them from seeking faculty jobs at all. Both grant proposal and peer review success rates can be higher for men than for women, because of implicit biases in the evaluations of the competence of women~\cite{katz:etal:2014,van2015gender}.  Technical disciplines, including computer science, often have a normative expectation of intellectual brilliance, and in these fields women are less likely than men to seek doctoral degrees~\cite{leslie2015expectations}. Experiments using name and gender variations on resumes have found that both male and female faculty members tend to rate male applicants as more competent, more hireable, and worthy of more mentoring than female applicants~\cite{moss2012science}. Taken together, it appears reasonable to expect strong and pervasive evidence of gender bias in faculty hiring outcomes across computer science.

Other studies have argued that the evidence of bias is lacking, even if it may have existed in the past. For instance, a review of 30 years of research on the leaky pipeline found that while gender differences were substantial prior to the 1990s in STEM fields, the gap has since closed~\cite{miller2015bachelor}. A separate review article surveyed literature on mathematical abilities in children, attitudes toward math-intensive fields, and access to, persistence in, and remuneration for faculty, concluding that no evidence of systematic gender bias exists today~\cite{ceci2014women}. One recent study controversially claimed to find a 2-to-1 preference for female applicants over male applicants in STEM tenure-track faculty positions, based on a hypothetical hiring scenario~\cite{williams2015national}. However, the experimental design did not include applicant publications, presentations, or reference letters, and thus it is unclear the degree to which these results reflect real preferences, aspirations, or political correctness. Even if the evidence is real, identifying its cause remains difficult. For instance, some studies argue that the critical variable underlying female underrepresentation is not gender itself but differences in personality \cite{correll2001} and structural position~\cite{xie98}; better access to resources for hiring, reviewing, and publishing~\cite{cole1984productivity,xie98,ceci2011understanding}; or the lower likelihood of workplace sexual harassment~\cite{ilies2003reported}, that happen to correlate with being male.

The role of gender in shaping outcomes in faculty hiring is difficult to assess, in part, because the hiring process itself is complicated and opaque. In real faculty searches, applicants will vary along dimensions of gender, productivity, subfield, doctoral prestige, postdoctoral experience, references, and more; applicants apply to many, but not all searches; and both applicants and institutions have internal, often undeclared preferences. Our aim in this paper is not to model all of these complexities. Instead, we adopt the more narrow goal of estimating the effective role of measurable factors like gender, productivity, and institutional prestige on observed faculty hiring outcomes. We do this by formulating a network model of the yearly matching process of applicants to faculty openings, which we parameterize to allow us to quantify the impact of different features of faculty applicants.
This approach allows us to investigate gender balance in the hiring histories of individual institutions and in individual faculty placement.

We begin by describing the faculty hiring and scholarly productivity data sets and the statistical features we derive from them. We then formulate a network model for faculty placement, check its accuracy in reproducing patterns found in the real hiring network, and use it to test a variety of hypotheses about the model features. Finally, we discuss our results in the context of other findings on gender inequality and highlight strengths and weaknesses of our analysis, before concluding. 

\begin{figure}[t]
	\centering
	\includegraphics[width=0.475\textwidth]{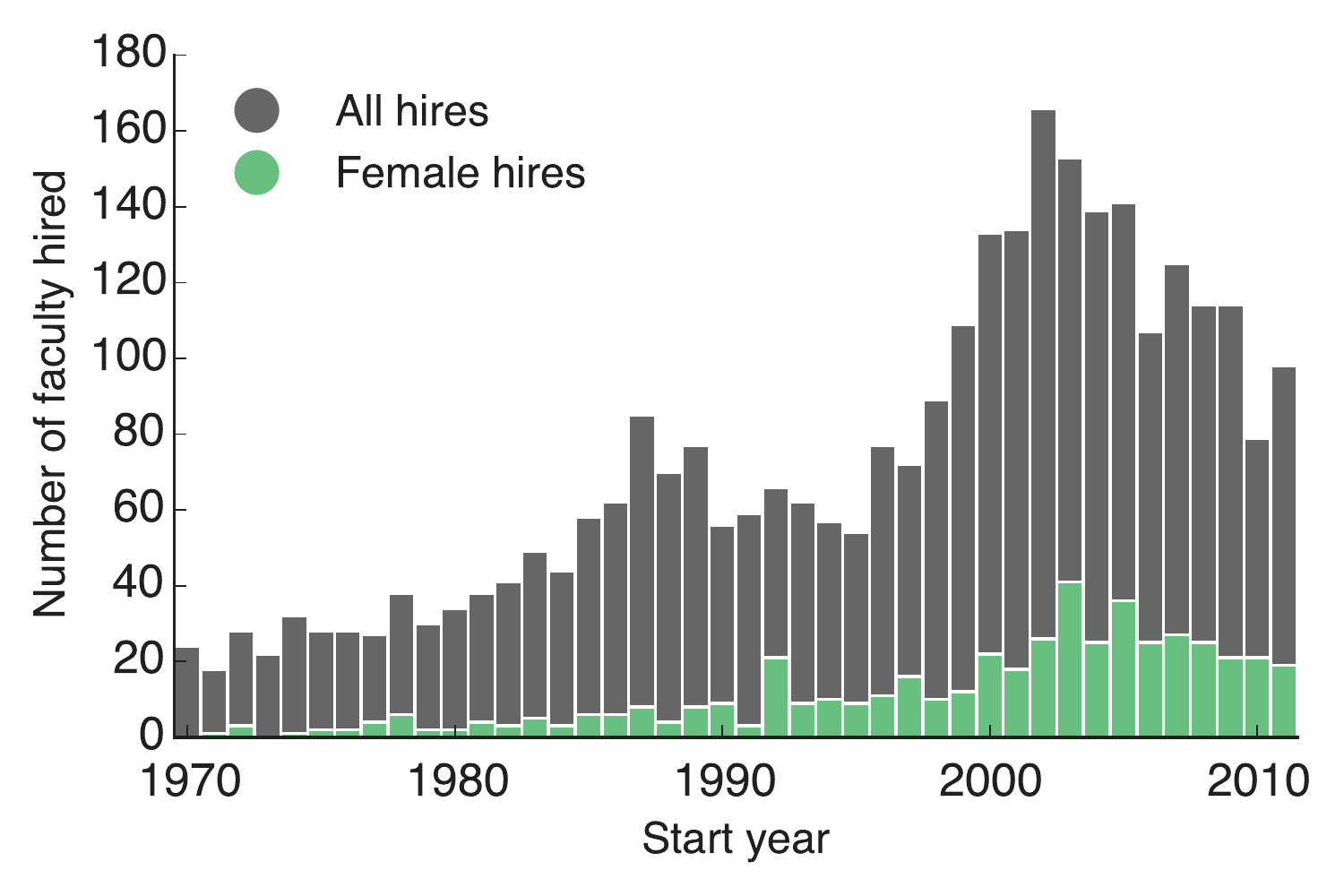}
    	\caption{For the 2659 computer science faculty in our sample (collected in 2011), the distribution of years in which they were first hired as an assistant professor.}
	\label{fig:basicstats}
\end{figure}

\section{Data and Features}

The primary data set that we used is a comprehensive, hand-curated list of the education and academic appointment histories of tenure-track or tenured computer science faculty~\cite{clauset2015systematic}. This data set covers the 205 departmental or school-level academic units on the Computer Research Association's authoritative Forsythe List of Ph.D.-granting departments in computing-related disciplines in the United States and Canada.\footnote{\url{http://archive.cra.org/reports/forsythe.html}} For each of these units, the data set provides a complete list of regular faculty from the 2011--2012 academic year, and for each of the 5032 faculty listed, it provides partial or complete information on their education and academic appointments, obtained from public online sources, mainly resum\'es and homepages.

Within this group, we selected the 2659 faculty who both received their Ph.D.\ from and held their first assistant professorship at one of these institutions, and for whom the year of that hire is known and occurred in 1970--2011. Figure~\ref{fig:basicstats} shows the distribution of these hire dates. The first requirement ensured that we modeled the relatively closed North American faculty market; roughly 87\% of computing faculty received their Ph.D.\ from one of the Forsythe institutions, and past analysis has shown that Canada and the United States are not distinct job markets in computer science~\cite{clauset2015systematic}. A number of faculty were removed in this step because the location of their first assistant professorship was not known; these were mainly senior faculty. The second requirement allowed us to extract a yearly time series of applicants and openings, and thus use a more realistic model of faculty hiring over time. Of the included faculty, women made up 16.1\%, which was not significantly different from the fraction in the discarded set ($p\!=\!0.92$, $\chi^{2}$), and the changes in institutional rank (see next subsection) were not significantly different between men and women in the discarded set ($p\!=\!0.325$, Mann-Whitney). Thus, our inclusion criteria are unlikely to bias our subsequent results.

We modeled the hiring process using a parametric model of edge formation in the faculty hiring network, in which the probability that a particular applicant is matched to a particular job opening depends on features of both applicant and opening. These features were (i) an applicant's gender, (ii) the prestige of the hiring institution, (iii) an applicant's scholarly productivity, (iv) an applicant's postdoctoral training, (v) the prestige difference between doctoral and hiring institution, and (vi) whether those institutions are in the same or different geographic regions. For each, we describe the way the feature was constructed and provide some simple statistics describing their relationship to gender.

\textbf{Institutional prestige.}
From the education and appointment data, we constructed a faculty hiring network, a directed multigraph where each node is an institution and each Ph.D.\ graduate from an institution $u$ who began as an assistant professor at $v$ is represented by a single directed edge $(u,v)$. Each node in this network is annotated with its institution's prestige rank~\cite{clauset2015systematic}, which is also given in the primary data set.

The prestige rank of an institution quantifies its ability to place its graduates as faculty at other prestigious institutions. Formally, $\textrm{rank}(u)$ is the mean rank of $u$ across all orderings that have the minimum number of ``violating'' arcs, i.e., an upward-pointing arc $(u,v)$, where $\textrm{rank}(v)$ is better than $\textrm{rank}(u)$. Such a ranking is called a \textit{minimum violation ranking} (MVR) and is a common way to measure prestige in social systems~\cite{devries:1998,henrich:gilwhite:2001}. The prestige ranking we used was obtained by sampling the MVRs for the full faculty hiring network, and it represents a hierarchy on the institutions in which only 12\% of edges violate the ranking, i.e., only 12\% of individuals were hired at an institution more prestigious than their doctorate institution. This ranking correlates with the popular but widely criticized~\cite{bastedo:bowman:2010} computer science ranking by \textit{U.S.\ News \& World Reports} ($r^{2}\!=\!0.80$), but it has the advantages of covering the complete Forsythe list and being based on the collective hiring decisions of the departments themselves.

We constructed two features using these ranks:\ the rank difference $\Delta \textrm{rank}(u,v)$ between the applicant's doctoral institution $u$ and the hiring institution $v$, and the $\textrm{rank}(v)$ of the hiring institution alone.

Comparing female and male faculty in our sample, we found no significant difference in the ranks of the doctoral institutions ($p\!=\!0.41$, Mann--Whitney) or the hiring institutions ($p\!=\!0.12$, Mann--Whitney). The distribution of the rank differences quantifies the degree to which applicants tend to move up or down the ranking when they take a faculty position (see Table~\ref{table:ranks}). We found no significant difference in the rank differences between men and women, both including ($p\!=\!0.33$, Mann--Whitney) and excluding ``self-hires" ($p\!=\!0.11$, Mann--Whitney), i.e., cases in which a university hires one of its own graduates. We did find a significant difference in the rates of self-hires, with 9.4\% of women being self-hired compared to 6.1\% of men ($p\!=\!0.02$, $\chi^{2}$). Altogether, men and women are trained and hired at similar rates across prestige rankings.

\begin{table}[h!]
\begin{center}
\begin{tabular}{r|cc}
	 & down & up  \\ \hline
	men & 1877 (79.3\%) & 491 (20.7\%) \\ 
	women & \hspace{0.5em}357 (81.0\%) & \hspace{0.5em}84 (19.0\%) 
\end{tabular}
\end{center}
\vspace{-3mm} 
\caption{Women and men move up in the prestige rankings at similar rates (excluding self-hires.)}
\label{table:ranks}
\end{table}

\textbf{Scholarly productivity.}
Publication records are an important factor in the evaluation of faculty candidates. For each applicant we assigned a feature that captures their scholarly productivity, controlling for subfield variations, prior to being hired into their first assistant professorship.

To construct this feature we first collected a complete publication profile for each faculty from DBLP, an online bibliographic database\footnote{\url{http://dblp.uni-trier.de/}} that, in late 2015, indexed over 3.1 million publications written by over 1.6 million authors, mainly computer scientists, using manual name disambiguation as necessary. Through this procedure, we obtained publication records, including titles and publication dates, for 2528 (95.1\%) faculty in our sample. The few individuals for whom we could not identify a DBLP profile were assumed to have no publications.

Publication records in DBLP include journal articles, conference papers (which, in computer science, are peer reviewed), as well as workshop papers (which often are not). The perceived value of different publication types, particular venues, or position in the author list varies by subfield, and we did not attempt to account for these differences here. Instead, we used the number of publications
that each faculty had published by one year after starting their assistant professorship, but normalized to control for publication rate variability across subfields.
To construct this normalization, we first aggregated the text contained in all the paper titles of a particular faculty's DBLP profile, a technique that is common in semantic analysis of short texts~\cite{hong2010empirical}. We then applied Latent Dirichlet Allocation~\cite{blei2003latent} to obtain 10 topics or subfield distributions over words, which together captured the total variation in words across all publication records. As a side effect, we also inferred for each faculty a probability distribution over subfields that characterizes their individual publication record. To verify that these distributions were reasonable, we manually inspected the most common words in each topic and found good agreement with classic subfields in computer science. Similarly, we verified that the inferred topic distributions for a set of well-known computer scientists aligned with their known specialities.

For each subfield, we computed a distribution over paper counts, weighted by each faculty's inferred emphasis on that subfield. For each faculty, we computed a single composite $z$-score for their overall productivity by taking a weighted average of $z$-scores over subfield distributions, with weights given by the faculty's subfield probability distribution. The result is a feature that represents each person's relative productivity, controlled for their own distribution of work across subfields and the norms within those subfields.

Productivity scores do not differ between men and women. This is true even when we consider only men and women who moved up the ranks and, separately, men and women who moved down ($p\!>\!0.05$, Mann--Whitney).  Median productivity scores for men and women in each of these categories are reported in Table~\ref{table:productivity}. We did find that individuals with postdoctoral experience have significantly higher productivity scores than individuals without postdoctoral experience ($p\!<\!0.01$, Mann--Whitney). This was true for men and women, separately and together. This is not surprising, as postdoctoral training allows more time to write papers prior to going on the faculty job market. As we note below, separate treatment of productivity and postdoctoral training allowed us to assess whether or not there is intrinsic value in postdoc experience beyond providing additional time to publish papers.

We note that the productivity scores of men and women do differ when we restrict our analysis to include men and women hired after 2002 (the median start year for women). Among these individuals, men are significantly more productive than women ($p\!=\!0.03$, Mann--Whitney). This finding supports the existence of a productivity gap in recent years, despite the previously mentioned studies, which suggest that such gaps have narrowed or closed over time in other disciplines \cite{arensbergen2012,xie98}.

\begin{table}[h!]
\begin{center}
	\begin{tabular}{r|ccc}
		& down & up & all \\ \hline
		men & -0.322 & -0.207 & -0.327  \\ 
		women & -0.331 & -0.215 & -0.329 
	\end{tabular}
\end{center}
\vspace{-3mm} \caption{Median $z$-scores by gender and by whether a faculty moved up or down the ranking for their faculty position. We find no significant differences comparing men and women's productivity scores in each of these categories. Median values are negative indicating that productivity scores are right-skewed due to prolific faculty.}
\label{table:productivity}
\end{table}

\textbf{Geography and postdoctoral training.}
Geography and postdoctoral training were captured in two binary features. For the former, we assigned a value of 1 if the pair $(u,v)$ spanned two institutions in the same geographic region (U.S.\ Census regions plus Canada), and a 0 otherwise. For the latter, we assigned a value of 1 if a person had any postdoctoral experience recorded in our primary data set, and a 0 otherwise.

We found no difference in the percentages of men and women graduating and being hired in the same geographic region ($p=0.12$, $\chi^{2}$). Of the people falling into this category, we next asked whether movement up or down in the ranks was linked to gender, and we found no evidence to suggest that these variables were related ($p\!=\!0.72$, $\chi^{2}$). We did find, however, that for individuals who changed geographic regions, men were significantly more likely than women to have moved up in rank ($p\!=\!0.01$, $\chi^{2}$). These results are presented in Table~\ref{table:geo}. Additionally, conditioned on moving up the ranks, men changed geographic regions significantly more than women ($p\!=\!0.03$, $\chi^{2}$), with 67.8\% of men changing regions compared to only 48.7\% of women. 

\begin{table}[h!]
\begin{center}
	\begin{tabular}{r|cc}
		& down & up \\ \hline
		men & 1150 (85.7\%) & 192 (14.3\%) \\ 
		women & \hspace{0.5em}220 (92.1\%) & 19 (7.9\%) 
	\end{tabular}
\end{center}
\vspace{-3mm} \caption{For individuals graduating and being hired in separate geographic regions, men are significantly more likely to be moving up the ranks ($p=0.01$, $\chi^{2}$).}
\label{table:geo}
\end{table}

We found that, in general, women were significantly more likely than men to have postdoctoral experience. 24.1\% of women in the dataset completed at least one postdoc compared to only 19.3\% of the men ($p\!=\!0.03$, $\chi^{2}$). Having postdoctoral experience, though, did not make women any more or less likely to move up the ranks than men ($p\!=\!0.92$, $\chi^{2}$), as displayed in Table~\ref{table:postdoc}.

\begin{table}[h!]
\begin{center}
\begin{tabular}{r|cc}
		 & down & up \\ \hline
		men & 347 (86.3\%) & 55 (13.7\%) \\
		women & \hspace{0.5em}80 (86.0\%) & 13 (14.0\%)  \\ 
	\end{tabular}
\end{center}
\vspace{-3mm} \caption{For individuals with postdoctoral experience, men and women move up the ranks at similar rates ($p=0.92$, $\chi^{2}$).}
\label{table:postdoc}
\end{table}

Finally, we note that the role of postdoctoral experience appears to have changed in recent years. Comparing individuals whose first assistant professorship began either before or after 2002, postdoctoral training rates were significantly higher following 2002, 28.1\% compared to only 15.5\% before 2002 ($p\!<\!0.01$, $\chi^{2}$). Men and women received postdoctoral training at similar rates post-2002, 29.5\% for women and 27.7\% for men ($p\!=\!0.68$, $\chi^{2}$), but the men who did were significantly more productive than the women ($p\!<\!0.01$, Mann--Whitney). We also note that after 2002 women \emph{with} postdoctoral training were not significantly more or less productive than men \emph{without} postdoctoral training ($p\!=\!0.44$, Mann--Whitney), suggesting that women faced additional obstacles which limited their productivity.

\begin{table*}[!htbp]
\centering
\begin{tabular}{l|c|ccc}
& & \multicolumn{3}{c}{model $f$} \\
 & observed & uniform & step & logistic \\ \hline
mean geodesic path length & \hspace{0.5em}2.23 & \hspace{0.5em}2.05 $\pm$ 0.01 & \hspace{0.5em}2.07 $\pm$ 0.01 & \hspace{0.5em}\textbf{2.16 $\pm$ 0.01} \\ \hline
mean local clustering coefficient & \hspace{0.5em}0.25 & \hspace{0.5em}0.34 $\pm$ 0.01 & \hspace{0.5em}0.38 $\pm$ 0.01 & \hspace{0.5em}\textbf{0.22 $\pm$ 0.01} \\ \hline
\% reciprocated hires & 18.95 & \textbf{14.52 \hspace{-0.1em}$\pm$ 0.81} & \hspace{0.5em}4.17 $\pm$ 0.34 & 13.93 $\pm$ 0.69 \\ \hline
\% reciprocating institutions & 14.25 & \textbf{13.23 \hspace{-0.1em}$\pm$ 0.77} & \hspace{0.5em}1.72 $\pm$ 0.21 & \hspace{0.5em}9.86 $\pm$ 0.61 \\ \hline
\% self-hires & \hspace{0.5em}6.62 & \hspace{0.5em}0.93 $\pm$ 0.18 & \hspace{0.5em}\textbf{3.74 $\pm$ 0.27} & \hspace{0.5em}1.95 $\pm$ 0.25 \\ \hline
\% placements within same region & 40.54 & 21.27 $\pm$ 0.77 & 24.48 $\pm$ 0.76 & \textbf{29.15\hspace{-0.1em} $\pm$ 0.75} \\
\end{tabular}
\caption{Network summary statistics used in model checking of uniform, step, and logistic choices of $f$. In each row, boldface indicates the model that best reproduces that characteristic of the observed network.}
\label{table:model:checking}
\end{table*}

\section{A model of the faculty market}

Faculty hiring is a complicated process, and the particular outcome of a faculty search can depend on a surprising variety of factors. 
Here, we aim to pare down this complexity to formulate a reasonably simple but still useful model of the faculty market as a whole in order to estimate the influence of different features on hiring outcomes in computer science. Our approach uses a data-driven statistical model of the observed outcomes and their features, which is distinct from models of strategic interactions among departments~\cite{kleinberg2015dynamic}.

We note two key properties of the faculty market: (i) assistant professor hires are made in rounds, generally once per year, and (ii) these hires are not independent of each other. This second property comes from the fact that two institutions cannot hire the same applicant. A faculty hiring network (where each directed edge $(u,v)$ represents the hiring a graduate of node $u$ as an assistant professor at node $v$) is thus the accumulation of yearly sets of such non-independent hiring edges.

We model this network assembly process by modeling the annual matching of candidates to openings in each year of the data. Systematic information on unsuccessful applicants and unfilled openings 
is not generally available for any year, and for this reason we make the simplifying assumption that matchings are made among the observed candidates and openings (the positions that were filled) in each year. This is not an unreasonable assumption:\ in practice, only a small fraction of faculty openings go unfilled each year, meaning that the set of successful applicants is a reasonable approximation of the top candidates across all searches.
Thus, for each year $t$, we first break the observed hiring edges $\{(u_{i},v)\}_{t}$, where $i$ indexes across all candidates, into two ``stub'' sets, one for the candidates $\{u_{i}\}_{t}$ and one for the openings $\{v\}_{t}$. We then generate a matching $M_{t}$ on these stubs using a probabilistic model $f$ that is parameterized by the pair-level features described in the previous section.

Regardless of the reasons why, in practice, hiring committees prefer applicants trained at more prestigious departments about 80\% of the time~\cite{clauset2015systematic}. 
We model this and other preferences of a typical hiring committee via a logistic function for the pairwise probabilistic model:
\begin{align}
f(\vec{x}[u_{i},v],\vec{w}) & \propto \left(1+\textrm{e}^{-\vec{x}[u_{i},v] \cdot \vec{w}}\right)^{-1} \enspace ,
\label{eq:logistic}
\end{align}
where $\vec{x}[u_{i},v]$ is a vector of features of the candidate-opening pair $u_{i},v$, and $\vec{w}$ is the global set of weights on those features that we learn from the data.

This choice of $f$ allows us to automatically capture two important special cases:\ if $f$ is independent of $\vec{x}$, then rank and other features play no role and the matching is equivalent to the popular configuration random graph model~\cite{molloy:reed:1995}; when $f$ is a step function on rank, and independent of other features, then hires are chosen uniformly at random from those trained at more prestigious departments, which is equivalent to the MVR ranking method used in~\cite{clauset2015systematic}. The step function is the simplest $f$ that depends on some of our features, and we use it as a baseline model later in order quantify the improvement from incorporating additional model features.

Applicants may also prefer openings at highly ranked departments, desiring the prestige and resources associated with these institutions. We model this preference by filling the openings $\{v\}_{t}$ sequentially, choosing an unfilled opening to fill with probability proportional to $1/\textrm{rank}(v)$ (where more highly ranked departments have smaller rank scores). Through this sequential matching process, our model fills each opening in a given year $t$ from the available candidates in that year.  Applying this process for each year $t$ from 1970 to 2011, the model assembles a full faculty hiring network. It is worth noting that this model is loosely similar to the popular exponential random graph model~\cite{robins:etal:2007}; however, in our formulation, edge formation is ordered and not independent, which requires a slightly different treatment.

We score the quality of our model by measuring its total error with respect to the observed placements, where total error is defined as the mean squared error (MSE) in the placements plus an L1 regularization term to prevent the model from overfitting. Mathematically,
\begin{align}
\textrm{err} = \frac{1}{m}\sum_{i=1}^{m} \left[\textrm{observed}(u_{i})-\textrm{model}(u_{i})\right]^{2} + \lambda\sum_{k}|\vec{w}_{k}| \, ,
\end{align}
where observed$(u_{i})$ is the observed placement rank of candidate $i$ and model$(u_{i})$ is the simulated placement rank. Using the MSE allows the model to receive partial credit for matching an applicant to an opening with rank similar to the observed rank, rather than, for example, receiving credit only if the applicant matches to the observed opening (which simply counts the number of correct placements). To estimate the model's parameters $\vec{w}$, we use a standard implementation of a direct search optimization algorithm (Nelder-Mead).	

\subsection{Model checking}

As a first step, we check that synthetic faculty hiring networks produced by our model have similar structural patterns to the observed network. We do this for each of three choices of $f$, the logistic function of Eq.~\eqref{eq:logistic} using all six features, as well as its two special cases, a uniform function and a step function. Using standard network summary statistics~\cite{Newman10}, such as the mean geodesic path length and the mean local clustering coefficient, as well as hiring-specific statistics on reciprocal hiring, self-hiring, and within-region placement, we compare the observed and simulated networks. Table~\ref{table:model:checking} summarizes the results of this exercise.

In general, we find very good agreement between the statistical properties of the real network and those generated by each of our models, with the logistic model performing best overall. Each of our models underestimates the rates of reciprocal hiring and self-hiring. This suggests that additional factors not present in our model likely influence these types of hires, perhaps related to the pre-existing social and professional connections associated with such hires.

Finally, we verify that the feature weights learned by our model are consistent under cross-validation in which sets of five randomly selected years of data are set aside for testing. Feature weights are largely stable across runs with only minor fluctuations that do not have a significant impact on modeling error.

\section{Results}

In the following sections, we examine gender's role in university faculty hiring at three levels by investigating (i) system-wide effects, (ii) hiring results for individual institutions, and (iii) hiring results for individual candidates. We conclude by forecasting when computer science will reach gender parity, should women's presence in the field continue to grow at the current rate.

\subsection{Market-level analysis}
\begin{figure}
\centering
\includegraphics[width=0.475\textwidth]{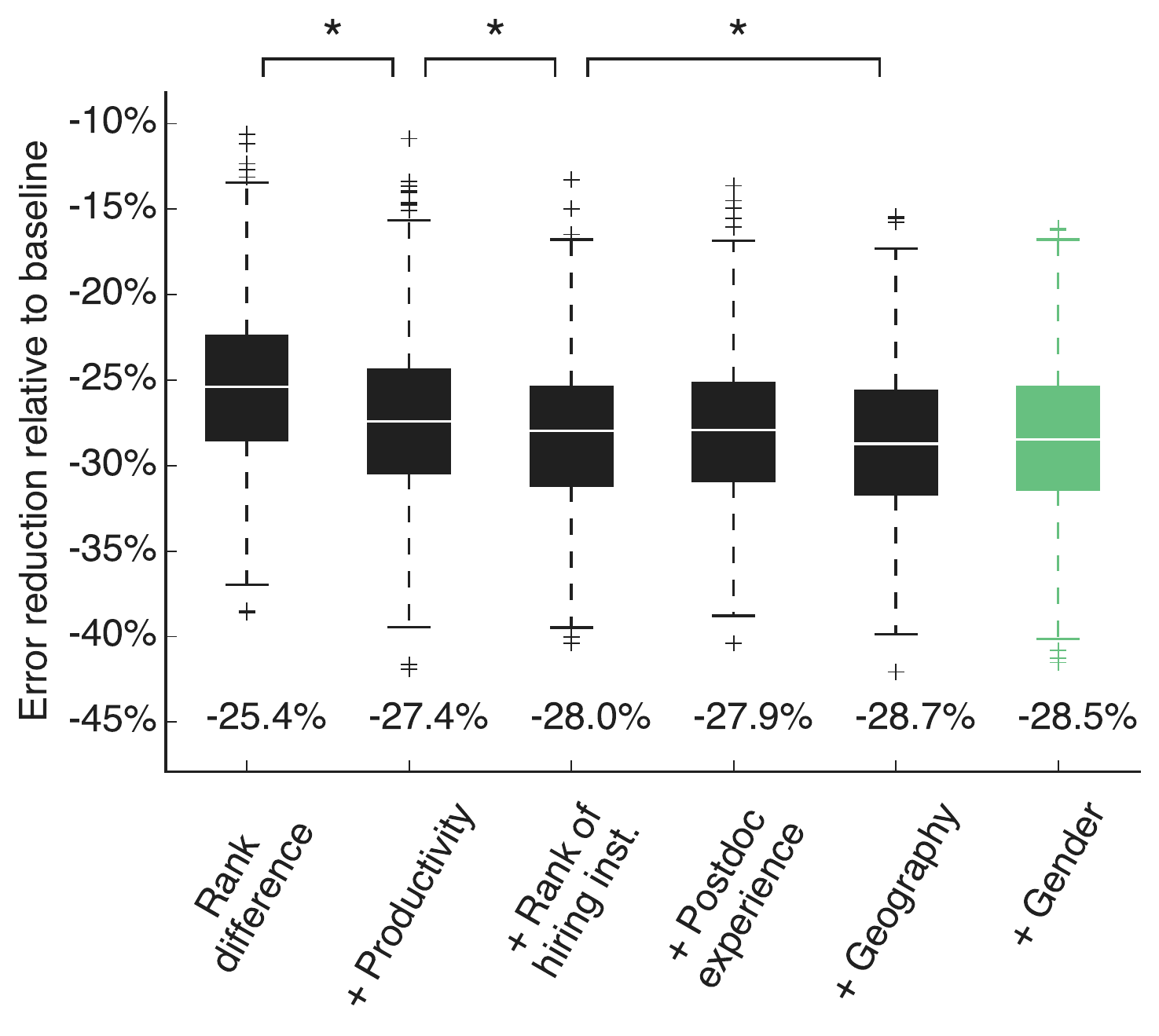}
\caption{Reduction of modeling error as features are added to the model. Percent reductions are computed relative to the step function model as a baseline. Median percent reductions are reported for each model, and attributes producing a significant reduction in error ($p\!<\!0.05$, Mann--Whitney) are marked with braces and asterisks.}
\label{fig:error}
\end{figure}

We trained a series of placement models by incorporating, one at a time, the attributes described in the previous section. The order in which attributes were added to the model was determined greedily: each remaining attribute was added separately to the previous model, and the attribute producing the greatest reduction of error was built into the subsequent model. Gender was incorporated last in order to determine if it significantly improved modeling results beyond the effects of all other variables. Figure~\ref{fig:error} shows the extent to which modeling error decreased as attributes were incrementally incorporated.

The list of attributes added to the model, in decreasing order of error reduction, was (i) rank difference between doctoral and hiring institutions, (ii) scholarly productivity, (iii) rank of hiring institution, (iv) postdoctoral training, and (v) whether doctoral and hiring institutions were in the same geographic region. It is perhaps unsurprising that rank difference and productivity yield the largest improvements in modeling results as these attributes are known to play key roles in faculty hiring. Incorporating the rank of the hiring institution also significantly improves modeling results ($p<0.01$, Mann--Whitney). Based on the sign of the inferred  coefficient, this suggests that the most prestigious universities are more selective in their hires and potentially value prestige more than lower-ranked universities.

Neither postdoctoral experience nor geographic information alone produced a significant change in modeling error. Together, however, these features accounted for a small but significant improvement. Because the productivity score had already been greedily added to the model prior to postdoctoral training, this result implies that postdoctoral training, in general, is only nominally useful beyond the extent to which it offers a trainee additional time to publish more papers and to thereby increase his or her productivity score. Geographic information, similarly, has little effect on modeling error. On its own, this finding suggests that issues of mobility do not strongly and systematically affect the placement of all faculty. We noted in Sec.\ 2, however, that men who moved up in the ranks are more likely than women who moved up to have changed geographic regions. Together, these findings suggest that mobility may play a small but real role in placement differences for some groups of men and women.

Finally, the addition of gender into the placement model did not significantly improve modeling results. We found this to be true both when we computed placement error for all faculty, and for women, separately. 
That the incorporation of gender does not significantly improve global error suggests that gender in and of itself does not systematically affect all hires beyond potential indirect effects encoded in other features, such as productivity. This finding echoes historical work \cite{cole1979}, which suggests that gender discrimination within science is not evenly distributed and warns that ignoring this non-uniformity risks promoting inequality.

That being said the weight assigned to gender was nevertheless non-zero, indicating that a subtle difference does exist. To convert this difference into more tangible terms, we calculated the number of additional papers a female candidate would need to publish in order to achieve the same job placement as an otherwise equivalent male candidate. Across subfields, on average, women must publish approximately one additional paper---a roughly 10\% increase in productivity---in order to compete on even footing with men.

\subsection{Institution-level analysis}
For faculty hiring to be free of uniform and systematic gender bias does not suggest that inequality cannot exist at the level of individual institutions. In this section, we explore this possibility directly by comparing the observed hiring at each institution with the distribution of outcomes drawn from our generative model of faculty placement. Using all features listed in previous sections, we simulated 1000 complete hiring histories, requiring as before that universities compete for candidates during each year of the process. For each simulation, we tracked the number of male and female hires by year and by institution, resulting in an evaluation of the gender balance of each department, taking into account the number of women on the job market when the department was hiring and the likelihood that those candidates would have been hired by the institution. The result is a set of institution-specific assessments that accommodate the non-independence of hires while controlling for placement likelihoods of candidates.

\begin{figure*}
	\centering
	\includegraphics[width=1\textwidth]{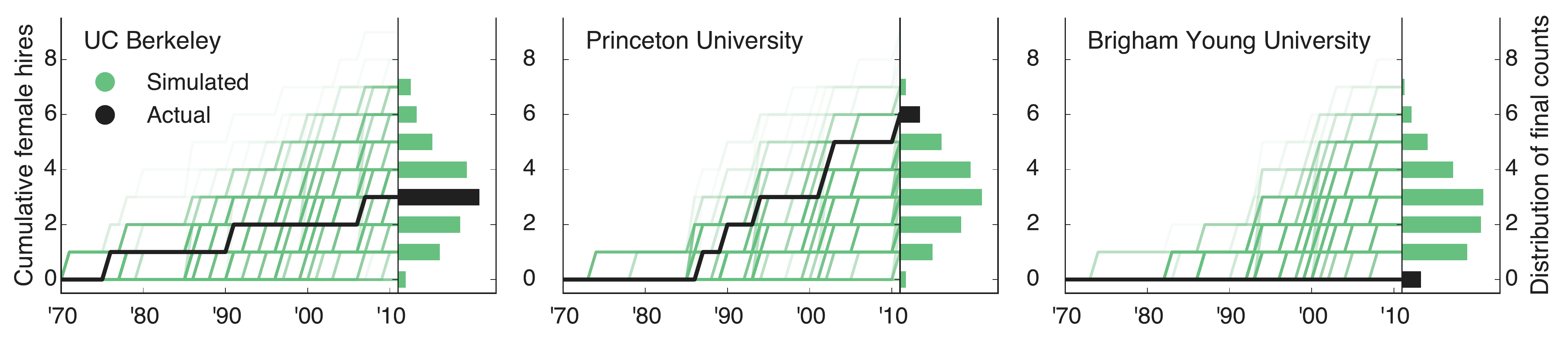}
	\caption{Three examples of model-based sampling of university-specific female hire distributions. Each green trajectory denotes the cumulative number of hires for a single simulation of the placement model at the indicated university. Running many simulations creates the distribution over final counts, shown on the right. The actual trajectory of hires made by the institution (within the data set) and the resulting final count are highlighted in black. UC Berkeley, Princeton, and Brigham Young represent examples of expected, female-skewed, and male-skewed hiring, as indicated by the location of the actual value within each sampled distribution.}
	\label{fig:trajectory}
\end{figure*}

In comparing each institution's actual number of female hires to the expected number under simulation, we find that most institutions perform very closely to their expected values. There are, however, institutions that exceed or fall short of the model's expectations. Figure~\ref{fig:trajectory} highlights universities in each of these three categories. 

\begin{figure}[!ht]
\centering
\includegraphics[width=0.475\textwidth]{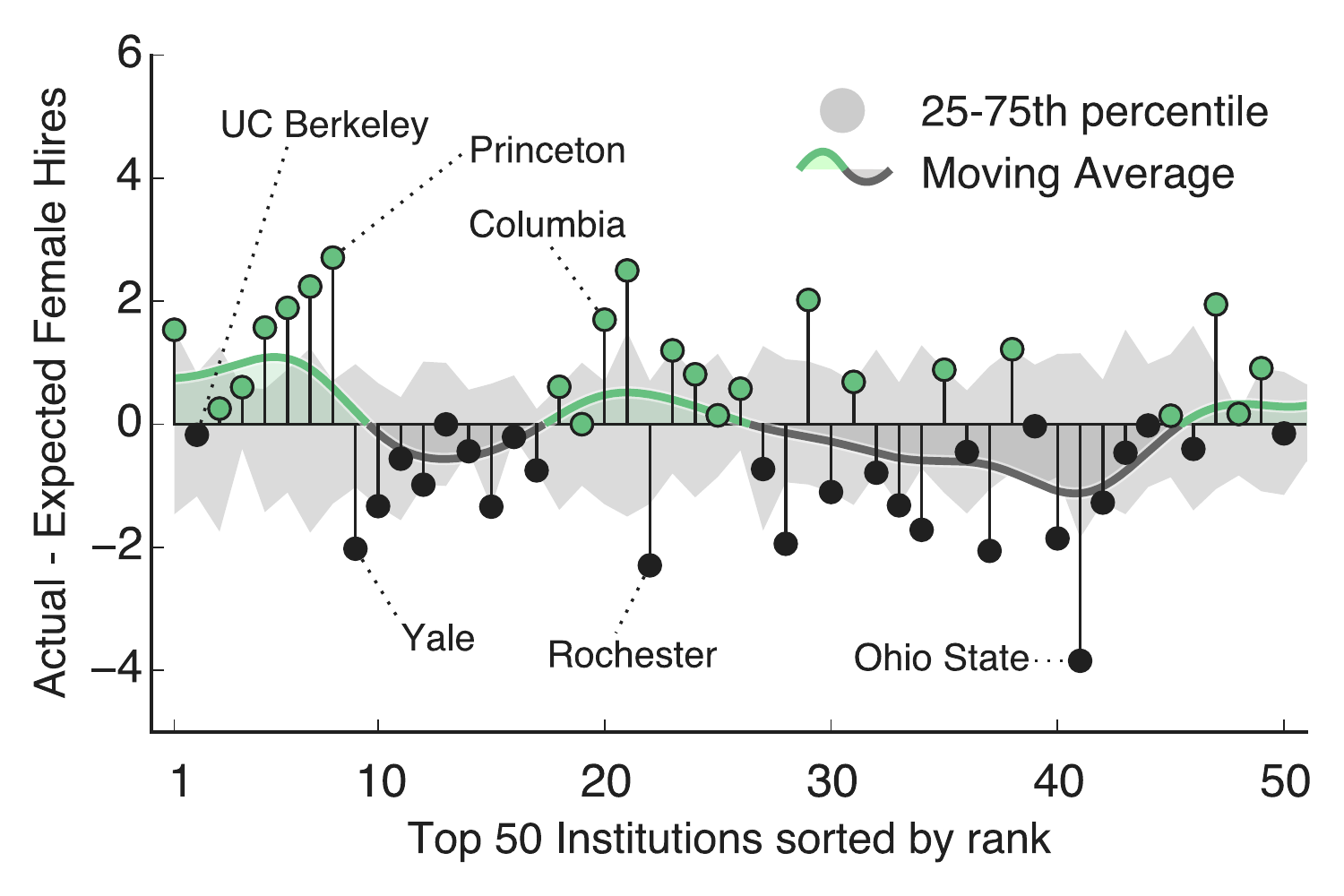}
\caption{Comparison of actual and expected female hiring over the top 50 institutions. Dots represent actual values minus expected values calculated from distributions samples as in Fig.~\ref{fig:trajectory}. The shaded region denotes the 25th-75th percentiles, based on modeling outcomes. Six particular universities are annotated. Top 10 schools hire slightly above expectations while ranks 11--20 hire below expectations. This suggests that the efforts by the highly-ranked schools to rectify any gender imbalance may have impeded the efforts of lower-ranked schools hoping to do the same.}
\label{fig:top50results}
\end{figure}

By comparing the results of many institutions, we asked whether female hiring patterns change as a function of rank. Figure~\ref{fig:top50results} illustrates the difference between actual and expected counts of women at the top 50 universities, sorted by rank. We note that top-ranked institutions (ranks 1--10) tend to hire more women than expected, while slightly lower-ranked institutions (ranks 11--20) typically hire fewer. This pattern may suggest that efforts made by top institutions to rectify instances of gender imbalance in their own departments could come at the expense of impeding similar efforts by lower-ranked institutions. 

\subsection{Candidate-level analysis}

\begin{figure}[!ht]
\centering
\includegraphics[width=0.475\textwidth]{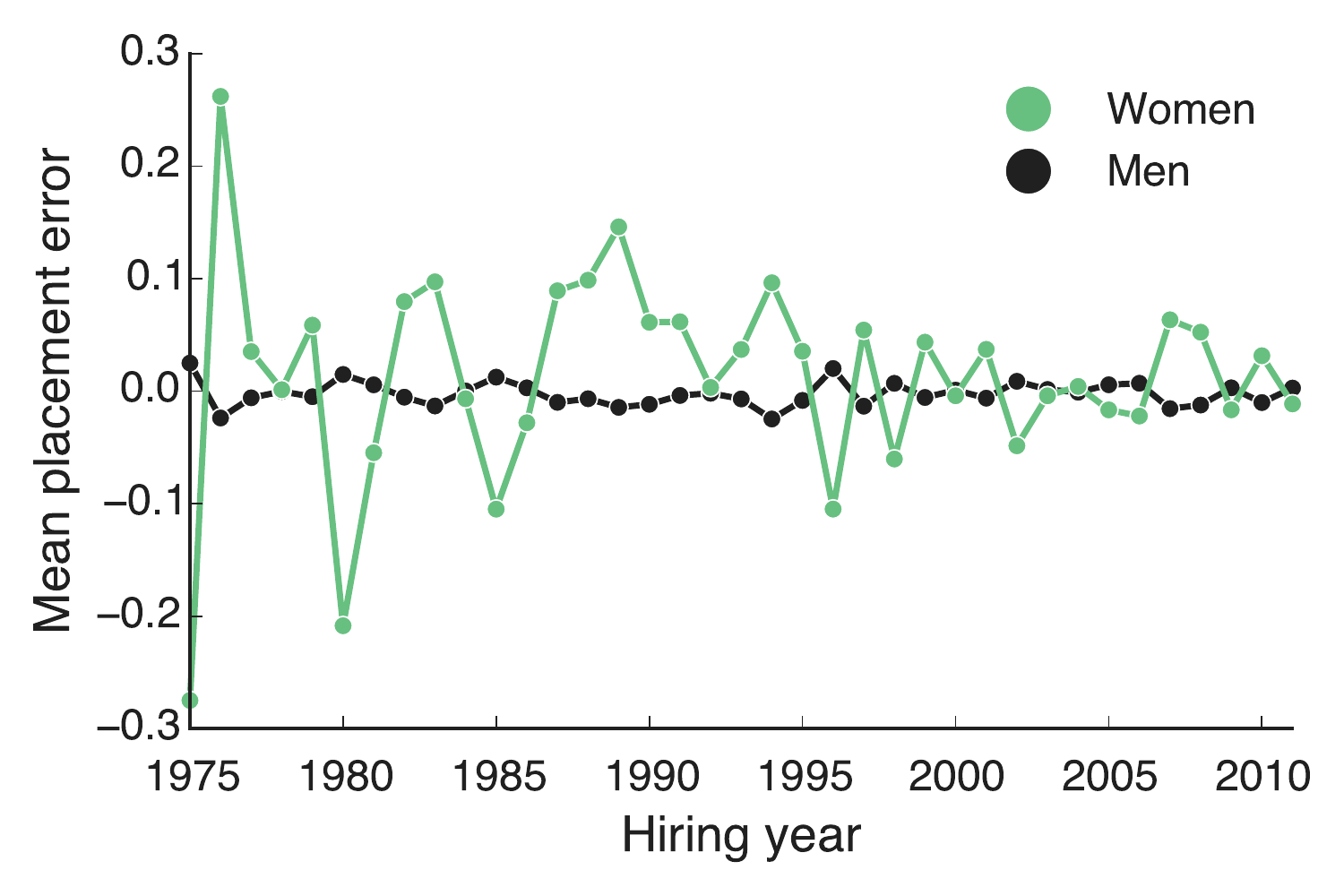}
\caption{Mean placement error by year. Placement error is computed as the difference between the rank of the institution where the person was hired and the rank of the institution where they placed under simulation. Higher variance in female placement error is within fluctuations expected due to lower female representation in the data set. Adjusted for yearly representation in the data, error is neither systematically increasing nor decreasing in time. }
\label{fig:errorvstime}
\end{figure}

Having analyzed faculty hiring at the system level and at the level of individual institutions in previous sections, we now investigate the placement of individual faculty. The complete simulations of the faculty market used in the institution-level analyses were re-analyzed for each individual faculty. Specifically, for each individual, we compiled a list of simulated placements and their frequencies, constituting a distribution of plausible outcomes for that person. By comparing the ranks of the institutions in an individual's list of plausible outcomes to that of their hiring institution, we obtained a distribution representing the amounts by which each person has over- or under-performed relative to their simulated outcomes. We separated these individuals by gender, and found that men and women meet or exceed model expectations at similar rates, though women are more likely to exceed expectations ($p\!<\!0.01$, Mann--Whitney). For under-performing individuals, however, men tend to fall short of their expectations by significantly larger amounts ($p\!<\!0.01$, Mann--Whitney).

We also find that individuals with postdoctoral training are more likely to outperform model expectations than those without this experience ($p\!<\!0.01$, $\chi^{2}$). This result is true for men and women, both separately and together, although women tend to exceed their expectations by larger amounts ($p\!<\!0.01$, Mann--Whitney). This implies that in the past, postdoctoral experience may have provided a strategic advantage to women looking to move up the ranks of the prestige rankings. With more men receiving postdoctoral training in recent years, however, it appears that what was once a competitive strategy may now be the norm.

Grouping individuals together by hiring year, we investigated how placement error is distributed over time. This allows us to assess the degree to which faculty hiring appears to have changed over the timeframe spanned by the dataset. Like the previous analysis, this is equivalent to looking at the average amount by which men and women over- or under-perform, collectively, in each hiring year. For instance, a pattern of women tending to under-perform early in the time period, and to over-perform later in the time period would be consistent with improved conditions for female faculty today. Instead, we see noisy, but relatively flat functions for the placement errors for both women and men (Fig.~\ref{fig:errorvstime}), with the difference in fluctuations by gender attributable to the difference in sample size. This pattern indicates that model errors in either direction are equally likely for men and for women, and for both recent hires and hires from several decades ago.

\begin{figure}[b!]
\centering
\includegraphics[width=0.475\textwidth]{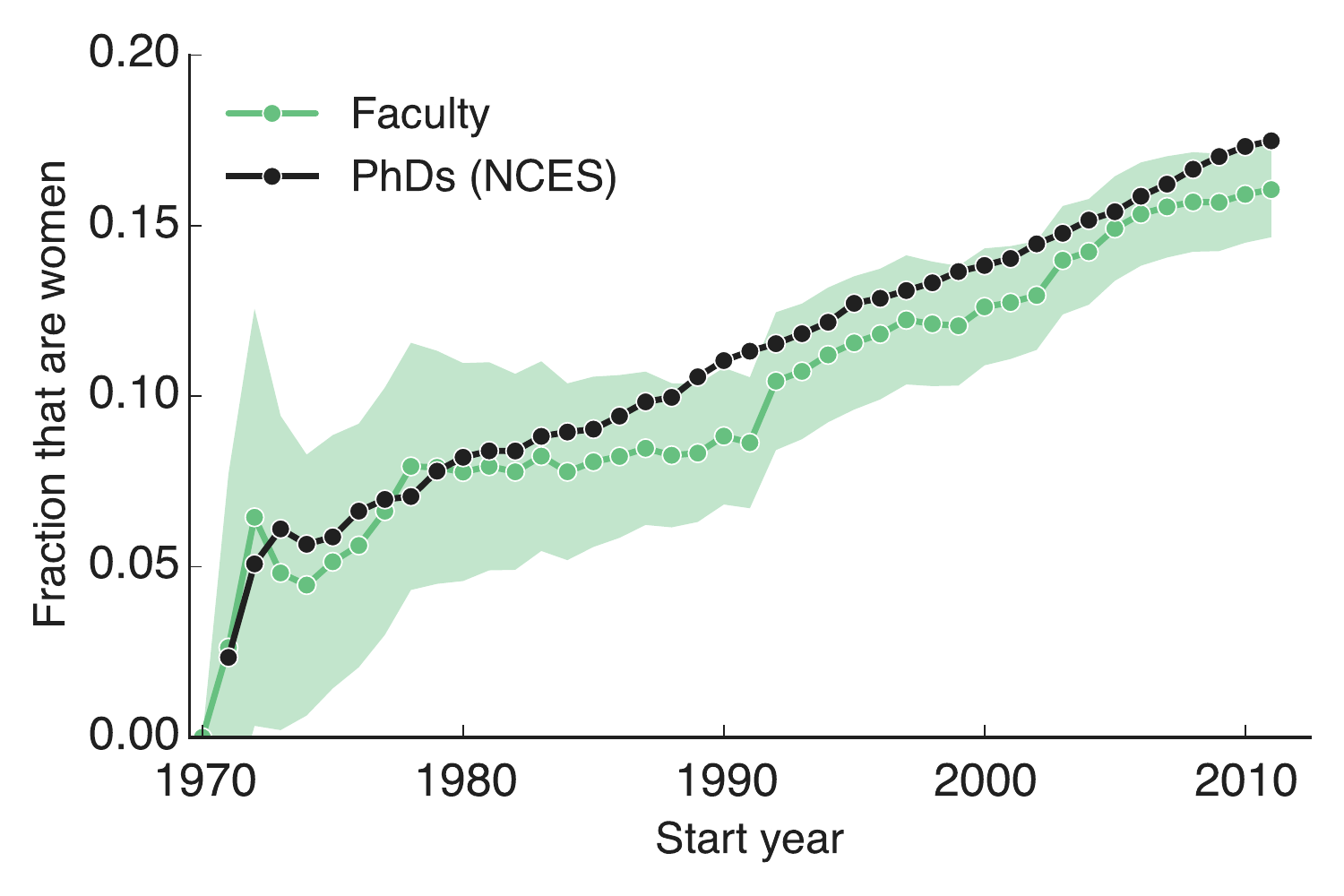}
\caption{Time series of the fraction of assistant professor hires since 1970 in our dataset that are women (green; with 95\% confidence intervals around the mean), and the fraction of computer science doctoral recipients since 1970 that are women (black). }
\label{fig:fraction_v_time}
\end{figure}

\begin{figure}[h!]
	\centering
	\includegraphics[width=0.475\textwidth]{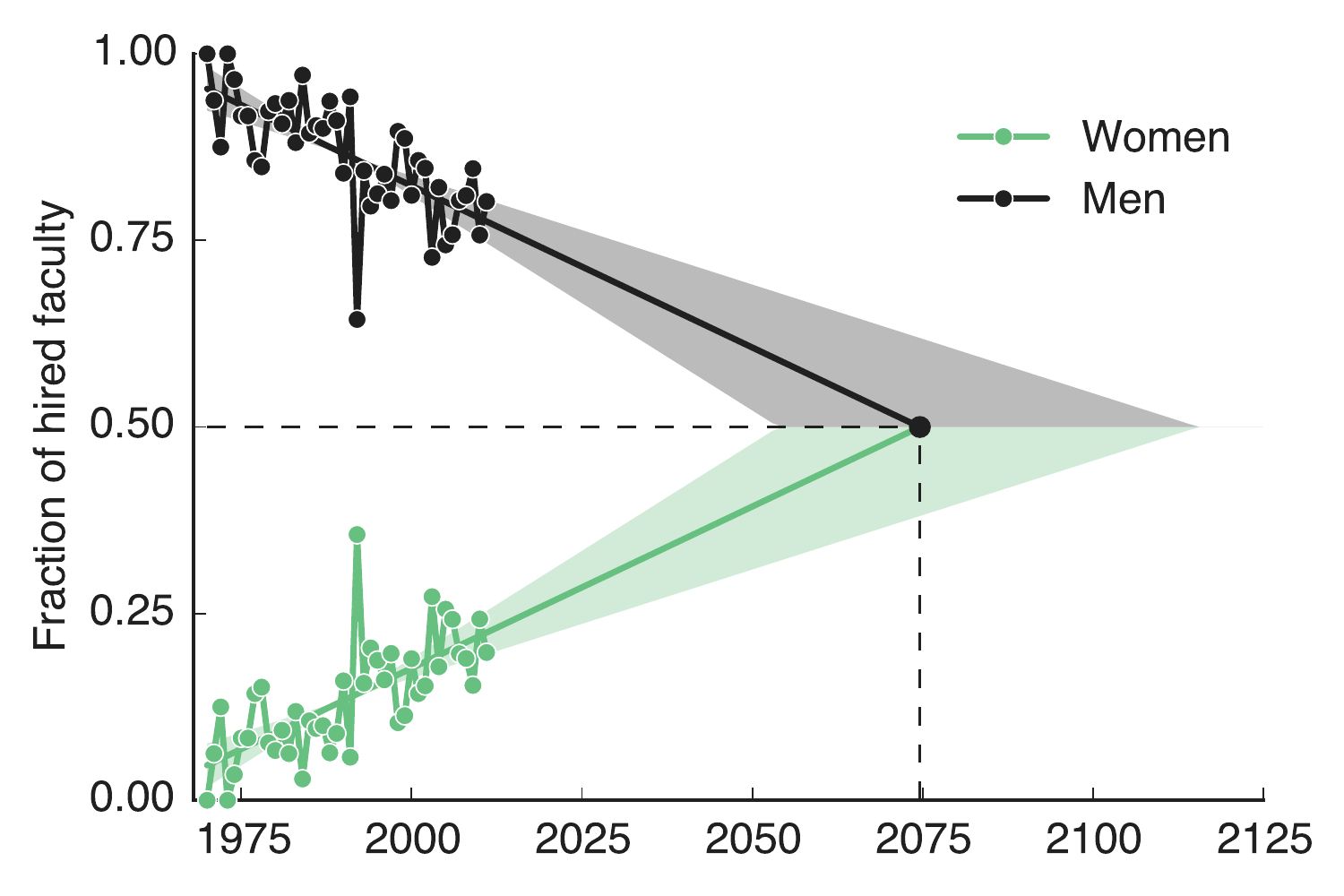}
	\caption{Gender ratio of assistant professors in computer science, by gender, and a projection for when gender parity will be reached. If the historical trend continues unaltered, gender parity will occur in approximately 2075. Shaded regions represent extrapolated 95\% confidence intervals from an ordinary least squares regression.}
	\label{fig:parity}
\end{figure}

\subsection{Long-term forecast for gender parity}
Over the four decades spanned by our data, the proportions of received doctoral degrees and assistant professor positions held in computer science by women have both steadily increased, from around 5\% to roughly 20\% (Fig.~\ref{fig:fraction_v_time}). However, the share of new faculty positions held by women is on average about 1\% lower than the share of doctorates, which reflects the well-documented leakiness of the academic training pipeline~\cite{hill2010so}. While not a large number in magnitude, a 1\% gap is a substantial proportional difference (about 7--20\%) given that the gender ratio is so heavily skewed toward men. 

Nevertheless, the long-term trend in computer science is toward gender parity. To estimate when women and men will hold equal shares of new faculty positions, we fitted a simple linear model to the historical trend and extrapolated it into the future (Fig.~\ref{fig:parity}). Under this model, the share of positions held by women increases by 0.43\% per year on average, meaning that it will take roughly 60 years from 2012 to reach parity at the assistant professor level, with a 95\% confidence interval of 30--100 years. Full gender parity across all levels of faculty should then occur 30--40 years later, when the first gender-parity cohort of assistant professors begins to retire. 

\section{Discussion}
Here, we used a unique data set on the hiring of assistant professors in computer science from 1970--2011 to measure the importance of six features of candidates on observed hiring outcomes. Among these, doctoral prestige and scholarly productivity play an outsized role, while gender alone does not appear to be a significant factor in the typical hiring decision. At face value, these findings are consistent with a system that is not overtly biased by a candidate's gender.

However, we also found evidence of (i)~unexpectedly gender imbalanced hiring patterns at individual institutions, (ii)~significant differences between genders in rates and the effects of publishing and postdoctoral training, (iii)~differences between men and women who move up the prestige ranking, and (iv)~evidence of that higher ranked institutions' success at hiring female faculty may be limiting similar efforts at marginally less highly ranked institutions. The apparent conflict between these two sets of findings about the same faculty market shows that the role of gender in faculty hiring is subtle and generally not well characterized by simple statistics or broad generalizations. Overall, our results suggest that the actual faculty hiring market in computer science is neither extremely dire for women~\cite{moss2012science} nor extremely favorable~\cite{williams2015national}.

Under our model, the inclusion of candidate gender did not significantly improve its ability to correctly place faculty overall. There are at least three plausible interpretations of this behavior. First, gender could be an irrelevant feature in faculty hiring. This interpretation is implausible because we also found that gender correlates with postdoctoral training, productivity, and geographic mobility, especially in the past 10 years. Second, the effect of gender may not be included realistically in the model. Evidently, a uniform penalty or advantage based on gender does not help reduce placement error rates, and so the gender feature received a weight near zero. Or third, the primary effects of gender on placement are already incorporated into the model through other features that correlate with gender.

This latter interpretation is particularly plausible. For assistant professors who started since 2002, productivity scores correlate with gender, with men being on average more productive than women with the same amount of training ($p\!<\!0.01$, Mann-Whitney). Moreover, the productivity of women \textit{with} postdoctoral training is not significantly different from men \textit{without} it ($p\!=\!0.44$, Mann-Whitney), and under our model, women need to be about 10\% more productive, on average, in order to place at equal rates as men. That is, productivity already encodes gender-based differences, making a separate gender variable in the model redundant. The origin of this productivity gap seems unlikely to be related to inherent differences in talent or effort, and may instead be related to differential access to resources and mentoring~\cite{ceci2011understanding}, greater rates of hostile work environments or sexual harassment~\cite{ilies2003reported}, differences in self-perceptions~\cite{chang2005dispelling}, or other gender-correlated factors. Additional research is needed to investigate these possibilities.

Our findings that support the existence of a gender-driven productivity gap in recent years are at odds with several studies indicating that such gaps have narrowed over time or perhaps closed altogether in other disciplines \cite{arensbergen2012,xie98}. These studies, however, examine the total number of publications and citations accumulated over one's entire career whereas we focus on an individual's publication record up until one year after being hired. Differences in productivity at this stage have been noted previously \cite{long1992measures} and are most relevant to our study of faculty hiring, as these differences likely influence hiring as well as tenure decisions and thus the individuals observed in our dataset. Indeed, we find that women are overrepresented on the low end of our productivity measure and publish fewer papers per year on average for the first several years of employment. A better understanding of the causes behind this lag in productivity would inform faculty evaluation procedures  and tenure policies, potentially improving retention of women at this career stage.

The productivity gap also suggests that postdoctoral training has been one way for women to compete on an equal basis with men in the faculty market. For faculty who started prior to 2002, the rate of postdoctoral training was indeed higher among women than men, which may reflect a compensatory adaptation to a biased system~\cite{wyche2008good}. Since 2002, however, these rates have equalized, meaning that in a typical faculty search today, men are likely to appear more productive, on average, than women. Institutional self-hiring, i.e., becoming faculty at one's doctoral institution, may reflect a separate kind of compensatory adaptation. Across 40 years, women have been hired by their doctoral institutions at a greater rate than men, and this difference has grown significantly since 2002. Determining the extent to which these patterns reflect strategic responses to a changing market would shed new light on the underlying market structure.

The long-term trend in the gender ratio in computer science faculty hiring is toward parity. The pace, however, is glacial, and we estimate that it will take roughly 60 years to reach. There are two main reasons to want to accelerate this trend:\ (i)~social justice and the provision of equal opportunities~\cite{dovidio:gaertner:1996,crosby:etal:2003}, and (ii)~increased scientific innovation, creativity, and productivity~\cite{bassett_jones:2005,page2008,catalyst:2013,kets2015challenging}. Achieving parity sooner, however, is likely to require novel and concerted efforts, as the faculty gender ratio correlates strongly with the doctoral gender ratio (Fig.~\ref{fig:fraction_v_time}), suggesting that relatively little has changed, fundamentally, over the past 40 years.

For an individual computer science department aiming to improve its faculty gender balance, the non-independence of hires poses a thorny problem. We observe a rank-dependent pattern indicating that more highly ranked departments tend to have better than expected rates of female faculty hiring and retention (Fig.~\ref{fig:top50results}), potentially at the expense of those departments ranked just below, e.g., ranks 1--10 vs.\ 11--19, and ranks 20--25 vs.\ 26--40. Even if all departments wished to hire more female faculty, the more highly ranked institutions will tend to have a competitive advantage in attracting \textit{any} candidates. Thus, if many departments are competing to hire a small number of female candidates, the lower-ranked departments will tend to lose out. Broadening the pool of female candidates is one solution to this problem, which a recent experimental study showed has a direct improvement on the gender ratio  among faculty hires~\cite{Smith10102015}.

Because the hiring network data set is a snapshot of regular faculty in the United States and Canada in the 2011--2012 academic year, it necessarily omits any information about faculty who left or retired from computer science prior to 2012, who were hired since 2012, or who were hired at the associate or full professor level during our study period, e.g., faculty who spent time in industry or who did their assistant professorship outside of computer science or outside the U.S.\ and Canada. As a result, hiring and retention are confounded in our analysis, and the current gender imbalance at some departments may be smaller than what we estimate.
Were information on these missing individuals to become available, our model could be used to study questions about the leaky pipeline, e.g., do certain institutions or groups of institutions contribute more or less to women leaving the pipeline, or to compare the dynamics of the new-hire market and the senior-hire market. Another limitation of this data set is that it does not include information on other faculty variables, such as their ethnicity, which can be particularly skewed, e.g., with African American faculty~\cite{allen:etal:2000}, socio-economic background, or nationality. These represent important directions for future research.

The productivity feature developed here could potentially be improved. For simplicity, we assigned all publications equal weight in our analysis, which favors quantity over quality. A better feature, however, would combine a candidate's scholarly record with an estimate of its scholarly quality and the author's level of contribution. However, such an extension would be highly non-trivial, in part because quality is difficult to measure accurately and automatically, across subfields. In fact, reliably assessing publication quality is hard even for humans, particularly when that contribution is interdisciplinary~\cite{lattuca2001creating}. An automated tool for doing so would have value both for the scientometrics and text mining communities as well as hiring committees.

In our model, we used a logistic function to score potential matchings between candidates and hiring institutions. Allowing this function to take a more complex form could improve the model's accuracy, either through the incorporation of interaction terms or by adopting a richer functional form in place of Eq.~\eqref{eq:logistic}. Though we do not explore these possibilities here, such modifications could enrich future analyses in this area and offer a source of flexibility for adapting our modeling framework to suit other applications.

Faculty hiring networks provide a powerful new tool for understanding the dynamics of academic disciplines, and for investigating the role of different factors in shaping academic careers.
The computer science hiring network reveals substantial evidence that gender inequality is present, subtle, and non-uniform. For predicting faculty placement, doctoral prestige and relative productivity appear to be the most important variables. However, the correlation between productivity and gender raises the questions of why, how the gap can be closed, and how our assessments can be informed by its underlying causes. Although the details are different, the computing industry has an equally large gender imbalance. Employing a similar approach to industrial hiring networks and productivity may shed new light on its underlying causes and the means to address it.

\section{Acknowledgments}
The authors thank Bailey Fosdick, Abigail Z.\ Jacobs, Winter Mason, and Jennifer Neville for helpful conversations, and the BioFrontiers Institute at the University of Colorado Boulder for computational resources. This work was supported in part by the Butcher Foundation.

%

\end{document}